\documentclass[PRB,twocolumn]{revtex4}
\usepackage{}
\usepackage{epsfig}
\usepackage{graphicx}
\usepackage{amsfonts,amssymb}
\usepackage{amsmath}
\usepackage{color}
\usepackage {times}
\definecolor{refcolor}{rgb}{1.0,0.0,0.0}
\usepackage{lipsum}
\usepackage[colorlinks,citecolor=blue,linkcolor=blue,anchorcolor=blue,filecolor=blue,urlcolor=blue]{hyperref}

\newcommand{\be}{\begin{equation}}
\newcommand{\ee}{\end{equation}}   
\newcommand{\bea}{\begin{eqnarray}}
\newcommand{\eea}{\end{eqnarray}}
\newcommand{\ba}{\begin{array}}
\newcommand{\ea}{\end{array}}

\newcommand{\q}{{\bf q}}
\renewcommand{\k}{{\bf k}}
\newcommand{\Q}{{\bf Q}}

\begin{document}

\title{Possible pairing states in the superconducting bilayer nickelate}

\author{Dheeraj Kumar Singh$^1$, Garima Goyal$^1$, Yunkyu Bang$^{2,3}$}
 
\affiliation{$^1$Department of Physics and Materials Science,
Thapar Institute of Engineering and Technology, Patiala 147004, Punjab, India}

\affiliation {$^2$Department of Physics, POSTECH, Pohang 790-784, Korea}
\affiliation {$^3$Asia Pacific Center for Theoretical Physics, Pohang, Gyeongbuk 790-784, Korea}

\date{\today}


\date{\today}
\begin{abstract}
We examine various possibilities for the pairing mechanisms in the recently discovered bilayer-nickelate superconductor within the Bardeen-Cooper-Schrieffer framework. Unlike earlier studies, where only a pure $d$-wave or sign-changing $s$-wave superconductivity instability was investigated, our study explores the possibilities of mixed-state superconducting instability such as the one involving both $d$- and sign-changing $s$-waves. While assuming that the superconductivity arises because of the magnetic correlations, we examine the nature of the superconducting gap function associated density of states with various possible magnetic correlation wavevectors arising out as a result of multiple pockets owing to the multiple orbitals and bilayer splitting. We also explore the effect of differences in the nature of Fermi surfaces suggested by various studies.
\end{abstract}

\maketitle

\section{Introduction}   
The discovery of unconventional superconductivity in the high-$T_c$ cuprates
has ever since stimulated a continuous search to seek similar systems that can exhibit a relatively enhanced superconducting-transition temperature~\cite{stewart, sigrist}. This has led to the uncovering of superconducting properties of iron-based pnictides and chalcogenides~\cite{kamihara, takahashi, chen, sefat, rotter, johnston}, and of nickelate very recently~\cite{yang, sun, li, li1, hepting, pickett}. In many aspects, nickelates have several similarities with cuprates~\cite{anisimov, zhang, botana, sakakibara, si, zhang3} and iron-based superconductors~\cite{ dagotto, takahashi1, zhang1, ying, zhang2}. However, the differences are more striking~\cite{karp, werner, gu}. One of the major differences is the absence of any concrete evidence in favor of long-range order in nickelates~\cite{jiang,zhang8}. In particular, the bilayer nickelate La$_3$Ni$_2$O$_7$, which, as yet, shows the highest superconducting-transition temperature $\sim$ 80K, is paramagnetic metal at ambient pressure. La$_3$Ni$_2$O$_7$ becomes weakly insulating when subjected to a pressure $\sim$ 1GPa~\cite{sun,liu2,zhang7,jiang1,wang}. Upon further increase of pressure, the resistivity of La$_3$Ni$_2$O$_7$ drops, and finally it becomes superconducting near $\sim$ 14GPa, and interestingly, continues to remain so if the pressure grows well beyond.

Unlike cuprates and infinite layer nickelate, Ni has a fractional oxidation state corresponding to $3d^{+7.5}$ as the outermost shell configuration in the bilayer La$_3$Ni$_2$O$_7$. Therefore, La$_3$Ni$_2$O$_7$ is unlikely to be found in the Mott insulating state~\cite{sun,zhang4}. At high pressure, both the $e_g$ orbitals, \textit{i.e.}, $d_{x^2-y^2}$ and $d_{3z^2-r^2}$ contribute at the Fermi level as revealed by the band-structure calculations~\cite{yang, zhang}. Because of the bilayer splitting, there are four bands dominated by the $e_g$ orbitals, out of which, effectively, three cross the Fermi level, one almost filled bonding band dominated by $d_{3z^2-r^2}$ orbitals~\cite{zhang5}. This leads to one nearly circular pocket around (0, 0), one larger and another smaller pocket around ($\pi, \pi$) in the Brillouin zone, which differs from the Fermi surfaces in bilayer manganites ~\cite{mannella, dks1, dks2}. A majority of features are  confirmed by a recent angle-resolved photoemission spectroscopy (ARPES)~\cite{yang1}. However, a recent study suggests the absence of the smaller pocket dominated by $d_{3z^2-r^2}$ orbital around M~\cite{ywang}.

A relatively high transition temperature in La$_3$Ni$_2$O$_7$ indicates a critical role of electronic correlations as also revealed in several theoretical works~\cite{lechermann, christiansson, cao,liu1,tian}. A strong Coulomb interaction may favor half-filled $d_{3z^2-r^2}$ orbital and localization of electrons. In the phases supported by a huge pressure, a large bilayer coupling exists, which, consequently, results into a large superexchange interaction through $d_{3z^2-r^2}$ orbital between the two layers~\cite{cao, zhang3, lu}. A larger superexchange coupling may lead to local spin singlets with large pairing energy~\cite{qin,oh,yang4,zhang9}.

Several approaches have been adopted to study the nature of pairing symmetry. These approaches include functional-renormalization group studies~\cite{yang}, use of linearized-gap equation with the Green's function incorporating the correlation effects within the fluctuation-exchange (FLEX) approximation~\cite{sakakibara}, tensor-network method~\cite{qu} etc. Their findings suggest $s^{\pm}$-wave~\cite{liu3} as the leading instability whereas there are other works which indicate $d$-wave as the dominant instability~\cite{ lechermann}. The nature of superconducting gap parameter can be probed via experimental probes such as nuclear-spin resonance~\cite{botzel,dai,hinkov,guidi,inosov}, quasiparticle interference~\cite{hofmann,hanaguri,hirschfeld,huang,dks3,sprau,dks4,rashid} etc.

In addition to the ambiguity prevailing with regard to the nature of superconducting gap function, there is no agreement between different studies with regard to the spin-fluctuation momenta playing a critical role in the pairing mechanism if the superconductivity is mediated by spin fluctuations~\cite{lechermann1}. Most probable spin-fluctuation momenta currently under intense discussion are ($\pi, 0$)~\cite{zhang6}, (0.75$\pi$, 0.75$\pi$)~\cite{yang}, (0.5$\pi$, 0.5$\pi$)~\cite{luo} etc. when three bands cross the Fermi level. The latter momentum is also suggested in a recent work according to which only two bands may cross the Fermi level, i.e., cuprate-like Fermi surface, however, with an additional nearly circular pocket around $\Gamma$~\cite{yang1,wang2}. The ambiguity over the dominant spin-fluctuation momenta may arise because of the multiplicity of the bands crossing the Fermi surface as well as the absence of strong nesting. In a multiorbital correlated system, the momentum dependence of correlation functions may be very sensitive to the bandstructure details, for instance, affected by the interlayer coupling, in case of bilayer nickelate.

In this paper, we examine the nature of superconducting gap parameter within the Bardeen-Cooper-Schrieffer (BCS) framework for a variety of possible paring wavevectors associated with magnetic fluctuations, which have been widely discussed for the recently discovered bilayer superconductors. Instead of comparing the free energies for different superconducting gap parameter for different pairing wavevectors, we focus on the properties of the superconducting gap resulting from the geometrical shape of the Fermi surface by using an effective three-band model, which reproduces the essential features of the Fermi surfaces as revealed by the band-structure calculations. It may be noted that the free energy associated with different types of superconducting gap parameters may differ only slightly because of the absence of a robust commensurate nesting vector as indicated in various studies and the interplay of multiple degrees of freedom including the electron-lattice coupling. The latter can arise due to the presence of more than one orbital and multiplicity of bands multiplied further by the interlayer coupling. In our approach, a relatively large value of gap obtained self-consistently for a given set of parameters does indicate a more stable superconducting state.
\section{Model and Method}
We consider a minimal phenomenological three-band model
for the bilayer nickelate in accordance with the fact that there are effectively three pockets in the Brillouin zone. The band-structure
calculations for bilayer nickelate at high pressure, when the superconductivity
appears, find that the Fermi surface (FS) consists of
two hole pockets around M, one larger and another smaller. In addition, there is also one electron pocket around $\Gamma$. However, a recent calculation suggests the absence of smaller pocket around M, which can easily be handled in our calculation by simple removing the band responsible for the pocket. The non-interacting part of the Hamiltonian is given by ~\cite{bang}
\begin{eqnarray}
H &=& \sum_{i \k \sigma} \varepsilon_{h_i} (\k) h^{\dag}_{i\k \sigma} h^{}_{i\k
\sigma} + \sum_{\k \sigma} \varepsilon_e (\k) e^{\dag}_{\k \sigma} e^{}_{\k
\sigma} \nonumber \\
& & +\sum_{i j \k \k^{'} \uparrow \downarrow} V(\k,\k^{'}) h^{\dag}_{i \k
\uparrow} h^{\dag}_{i-\k \downarrow}
h^{}_{j\k^{'} \downarrow}h^{}_{j-\k^{'} \uparrow} \nonumber \\
& & +\sum_{\k \k^{'} \uparrow \downarrow} V(\k,\k^{'}) e^{\dag}_{\k
\uparrow} e^{\dag}_{-\k \downarrow} e^{}_{\k^{'} \downarrow}e^{}_{-\k^{'}
\uparrow} \nonumber \\
& & + \sum_{i \k \k^{'} \uparrow \downarrow} V (\k,\k^{'}) h^{\dag}_{i \k
\uparrow} h^{\dag}_{i -\k \downarrow} e^{}_{\k^{'} \downarrow}e^{}_{-\k^{'}
\uparrow} \nonumber \\
& & +\sum_{i \k \k^{'} \uparrow \downarrow} V(\k,\k^{'}) e^{\dag}_{\k
\uparrow} e^{\dag}_{-\k \downarrow} h^{}_{i \k^{'} \downarrow}h^{}_{i-\k^{'}
\uparrow}.
\end{eqnarray}
$h^{\dag}_{i\k \sigma}$ and $e^{\dag}_{\k \sigma}$ are electron creation operators in the $i$-th hole band and the electron band, respectively. $i, j$ = 1 and 2, for the larger and smaller hole pockets, respectively. The band dispersions are $\varepsilon_{h_1} (\k) =  t_1\varepsilon_1(\k)+t_2\varepsilon_2(\k)+ t_3\varepsilon_3(\k) + \epsilon_1$, $ \varepsilon_{h_2} (\k) = t_1\varepsilon_1(\k)+t_2\varepsilon_2(\k)+t_3\varepsilon_3(\k) + \epsilon_2$, and $\varepsilon_{e} (\k) = -t_1\varepsilon_1(\k) + t_2\varepsilon_2(\k)+t_3\varepsilon_3(\k) + \epsilon_3 $. Here, we have defined the followings: $\varepsilon_1(\k) = \cos k_x +\cos k_y$, $\varepsilon_2(\k) = \cos k_x \cos k_y$, and $\varepsilon_3(\k) = \cos  2k_x +\cos 2k_y$. The dispersion is defined in a way to keep the number of independent hopping parameters as minimal as possible without compromising with the salient features of the bands found through the band-structure calculations. In this paper, band parameters are set to be $t_1 = -1.0$, $t_2 = 0.65 $, $t_3 = -0.14$, $\epsilon_1 = 0.6$, $\epsilon_2 = -1.3$, and $\epsilon_3 =  -1.2$ in the unit of eV.

We assume a phenomenological form of the pairing interaction as
\begin{equation}
V(\k,\k^{'}) = V_M \frac{\kappa^2}{|({\k}-{\k^{'}})-{\Q}|^2
+\kappa^2}.
\end{equation}
The pairing interaction is repulsive in the momentum space and represents a short-range spin fluctuations. The interaction has a peak when the difference between the momenta $\k$ and $\k^{'}$ is equal to $\Q$. Information about the pairing momentum $\Q$ is often inferred from nature of symmetry breaking in immediate vicinity of the superconducting state when the concentration of charge carrier is modified just like in the high-$T_c$ cuprates or in the iron-based superconductors. For the bilayer nickelate, any concrete evidence for such a symmetry breaking in the so-called weakly insulating state is still awaited. However, according to the different band-structure calculations, several pairing momenta $\Q$s are under consideration. Below, we examine various possible scenario and supported pairing. Another parameter $\kappa$ is associated with the correlation length with the short-range magnetic fluctuations with wavevector $\Q$ where $\xi_{\Q}$ = $2\pi a/\kappa$, where $a$ is the lattice parameter.

Using the Bardeen-Cooper-Schrieffer (BCS) approximation, the superconducting (SC) order parameters (OP) along the two hole pockets and one electron pocket are given by
\begin{eqnarray}
\Delta_{h_i} (\k) & = &  \sum_{\k^{'} } V(\k,\k^{'}) \langle h_{ i \k^{'} \downarrow} h_{i -\k^{'} \uparrow}\rangle , \\
\Delta_{e} (\k) & = &  \sum_{\k^{'} } V(\k,\k^{'}) \langle e_{\k^{'}
\downarrow} e_{-\k^{'} \uparrow}\rangle.
\end{eqnarray}
The subscript $i = 1, 2$ for the two hole bands. After meanfield decoupling of the interaction terms in Eq. (1) and then using the definition of SC OPs given by Eqs. (3) and (4), one obtains
\begin{eqnarray}
\Delta_{h_i} (\k) & = &  -  \sum_{ j \k^{'} } [V_{h_i h_j}
(\k,\k^{'}) \Delta_{h_j} (\k^{'})  \chi_{h_j} (\k^{'}) \nonumber\\ & +& V_{h_i e}
(\k,\k^{'})\Delta_e (\k^{'})  \chi_e (\k^{'})],
\end{eqnarray}
\begin{eqnarray}
\Delta_e (\k)   &=&    -  \sum_{i \k^{'} } [V_{e h_i}
(\k,\k^{'}) \Delta_{h_i} (\k^{'})  \chi_{h_i} (\k^{'}) \nonumber\\ &+& V_{ee}
(\k,\k^{'})\Delta_e (\k^{'})  \chi_e (\k^{'})].
\end{eqnarray}
\noindent $V_{h_i h_j} (\k,\k^{'})$, $V_{h_i e} (\k,\k^{'})$, etc. are
the interaction as defined by Eq. (2) and $V_{h_i h_j} (\k,\k^{'})$ =$V
(\k_{h_i},\k^{'} _{h_j})$, $V_{h_i e} (\k,\k^{'})$ =$V (\k_{h_i} ,\k^{'}_e)$, etc. $\k_{h_i}$ and $\k_e$ are the momenta located along different hole and electron pockets, respectively. The pair susceptibilities $\chi_{h_i/e}(\k)$ are given by
\begin{eqnarray}
\chi_{h_i/e}(\k) &=& N(0)_{h_i/e} \int _0 ^{\omega_{c}} d \xi
\frac{\tanh (\frac{E_{h_i/e}(\k)}{2 T})}{E_{h_i/e} (\k)}.
\end{eqnarray}
\noindent $E_{h_i/e} (\k) =\sqrt{\xi^2 + \Delta_{h_i/ e}^2 (\k)}$ is the quasiparticle dispersion and $N(0)_{h_i/e}$ is the density of states (DOS) along different hole and electron pockets at the Fermi level, respectively. $\omega_{c}$ is
the cutoff energy of the pairing potential $V(\q)$.

\section {Gap solutions}
Unlike the iron-based superconductors, where the interpocket pairing process was particularly dominant because the size of Fermi pockets was much smaller than the magnitude of $\Q$, the situation is entirely different for the bilayer nickelate largely due to the presence of multiple Fermi pockets with one of them being significantly large. For this reason, we incorporate both the interband and intraband processes in our calculations. However, we also present the results when only the interband pairing is considered, in order to highlight the role of intraband pairing. This is particularly important as the larger hole pocket is of similar shape and size as the one and only one found in the single-layer cuprates.

Throughout the paper, we fix the
parameters $\kappa=0.1 \pi$ ($ \xi_{\Q} \sim 20a$),
$\omega_{AFM} =20$ meV, and $V_M=10$ eV. The choice of the band parameters mentioned earlier yields $N_{h_1}(0) = 0.43/$eV, $N_{h_2}(0) = 0.18/$eV, $N_e(0) = 0.22/$eV
so that $N_{h_1}(0)/N_e(0) \approx N_{h_2}(0)/N_e(0)
\approx 2.0$. An important difference from the iron-based superconductor is that the argument if $N_h(0) > N_e(0)$ then $|\Delta_h(k)| < |\Delta_e(k)|$ is not applicable as we shall see below. It is only for  the purpose of demonstration that we have chosen an arbitrary value of the pairing strength $V_M = 10$eV, which affects the magnitude of SC OPs and not their nature such as symmetry etc.
\section{Results and Discussion}
\begin{figure}
\hspace{-3.5mm}
   \includegraphics[scale=1.0, width=8.8cm]{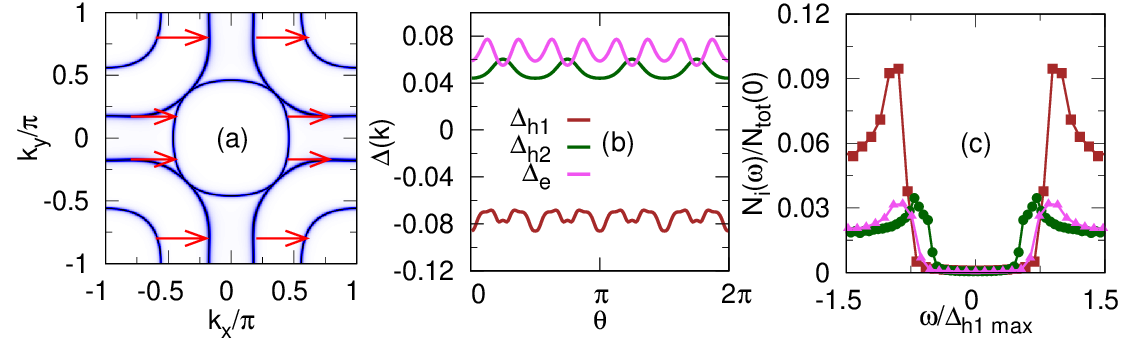}
    \vspace{-5mm}
   \caption{ (a) Nesting vector, (b) SC gap structure, and (c) DOS for the self-consistently obtained solutions for the pairing momentum $\Q \sim$ (0.5$\pi$, 0). The solution exists only  when the interband pairing is considered. }
    \label{1}
    \end{figure}

\begin{figure}
   \includegraphics[scale=1.0, width=9cm]{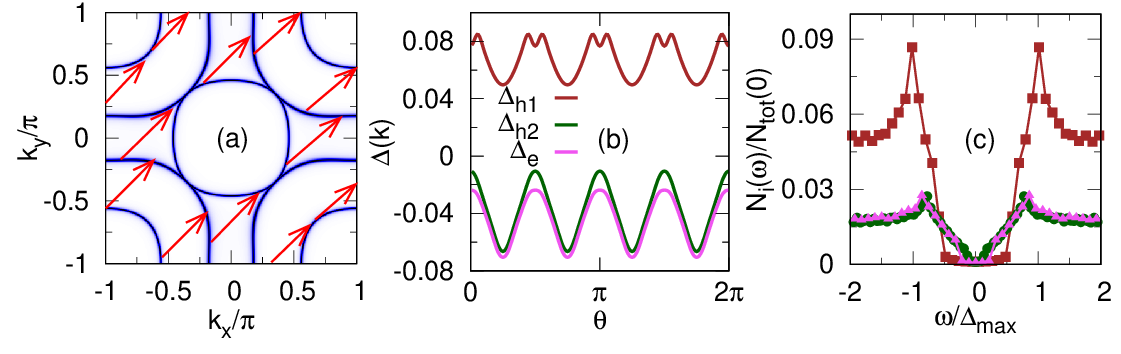}
    \vspace{-5mm}
   \caption{ (a) Nesting vector, (b) SC gap structure, and (c) DOS for the self-consistently obtained solutions for pairing momentum $\Q \sim$ (0.5$\pi$, 0.5$\pi$), when only the  interband pairing is considered. A large oscillation in gaps along the Fermi surfaces may be noticed while the gap sign on the larger hole pocket is opposite to the gap sign on the smaller hole pocket and electron pocket. }
    \label{2}
    \end{figure}
Figure~\ref{1} (a) shows the Fermi surfaces obtained within the three-band effective model. There are three pockets, one large and one small hole pocket around ($\pi, \pi$) whereas one electron pocket around $\Gamma$ in accordance with the first-principle bandstructure results reported in various works. There are several possible nesting vectors, which may potentially give rise to magnetic correlations. Here, we focus mostly on those which are commensurate except $\Q \sim$ ($0.75\pi$, $0.75\pi$).

One of the nesting vectors is $\Q \sim$$ (0.5\pi, 0)$ in between the nearly straight and vertical sections of the larger and smaller hole pockets (Fig.~\ref{1}). The nesting vectors between the nearly straight legs of the larger hole pocket is, however, relatively smaller in magnitude.
A self-consistently obtained superconducting gap function for the hole- and electron-pockets are shown in Fig.~\ref{1}(b), when only the interband pairing is considered. The sign-changing $s^{\pm}$-wave SC state is stabilized in such a way that the SC gap along the larger hole pocket has a sign opposite to that along the remaining two pockets. The average gap size is almost similar for the electron and the hole pockets, however, the nature of oscillation in the SC gap differs slightly for the different pockets. As expected, the DOS in the SC state has a $s$-wave like gap. It can be noted that the amplitude of oscillations in the SC gap is relatively larger when compared with the same for the iron-based superconductors, which can result from the large Fermi surfaces and a weak nesting. As we will see below that this feature remains persistent. A self-consistent solution could not be obtained if the intraband pairing is also included. The existence of self-consistent $s$- or $d$-wave or mixed-state solution only in the case of interband pairing is a signature of the important role of bandstructure in this class of superconductor.

The contribution for nesting along the horizontal or vertical direction with $\Q \sim$ $(0.5\pi, 0)$ wavevector comes only from two nearly straight sections of the hole pockets. On the other hand, the nesting is expected to be improved in the diagonal direction with the wavevector  $\Q \sim (0.5\pi, 0.5\pi)$ as all the nearly straight sections can contribute (Fig.~\ref{2}(a)). However, this is not well reflected in the average gap size, as the latter is not larger than the one obtained in the case of $\Q \sim$ $(0.5\pi, 0)$, which may result from the difference in the actual nesting vector and $(0.5\pi, 0.5\pi)$ (Fig.~\ref{2}(b)). Figure~\ref{2} shows various interband nesting vectors, gap functions, and DOS, when only the interband pairing is considered. As in $ (0.5\pi, 0)$ case, the larger hole pocket has opposite sign of SC gap in comparison to other two pockets. However, the amplitude of SC gap oscillation along all the pockets has increased comparatively. The gap at some points of the smaller hole pocket or electron pocket even approaches zero. As a consequence, the gap in the DOS has a mixed character$-$the high-energy portion of gap in the DOS shows a structure close to V-shape for all the pockets, which is more pronounced for the smaller hole pocket and electron pocket, while the low-energy behavior is similar to the $s$-wave SC state.

\begin{figure}
   \includegraphics[scale=1.0, width=8.8cm]{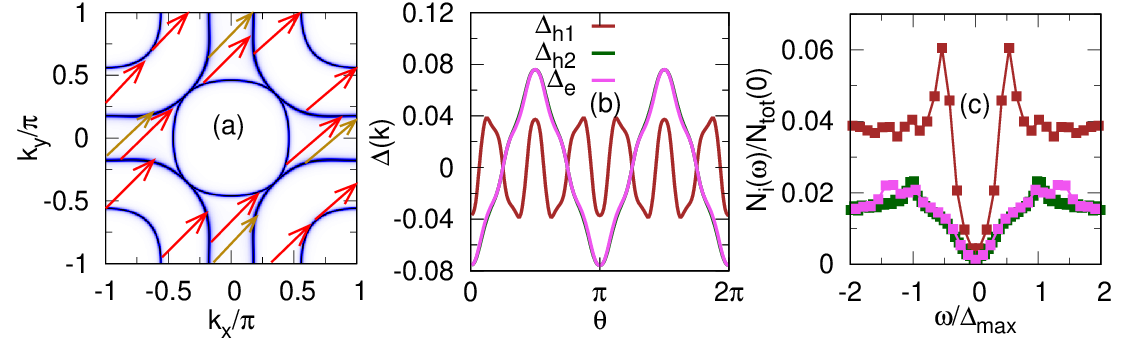}
    \vspace{-5mm}
   \caption{ (a) Nesting vector, (b) SC gap structure, and (c) DOS for the self-consistently obtained solutions for pairing momentum $\Q \sim$ (0.5$\pi$, 0.5$\pi$), when both the interband and intraband pairings are incorporated into the calculations. The SC gap has a $d$-wave like structure along the smaller hole pocket and electron pocket whereas a $d$-wave like shape, but the frequency of oscillations is doubled because another maxima of same strength appears between two maxima. }
    \label{3}
    \end{figure}

\begin{figure}
   \includegraphics[scale=1.0, width=9cm]{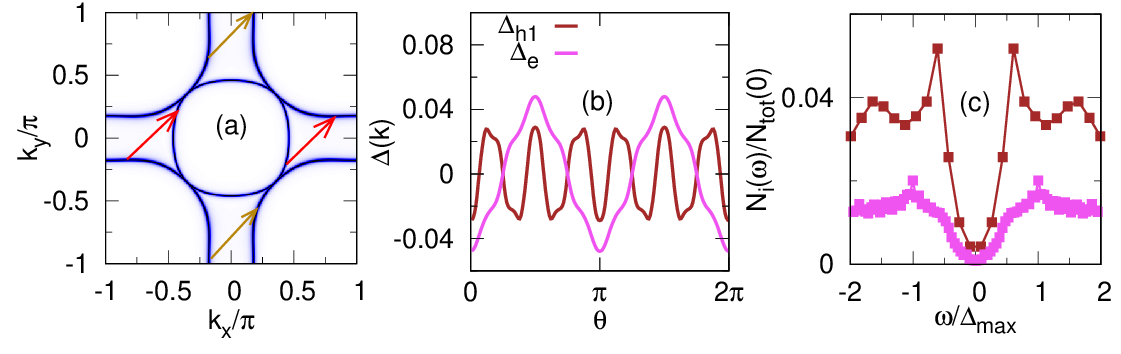}
    \vspace{-5mm}
   \caption{(a) Nesting vector, (b) SC gap structure, and (c) DOS for the self-consistently obtained solutions for pairing momentum $\Q \sim $ (0.5$\pi$, 0.5$\pi$) when the smaller hole pocket is ignored as suggested in a recent study~\cite{wang2}. Both the intraband and interband pairings are incorporated. The SC gaps have similar structures as in Fig.~\ref{3} except that the size is reduced. }
    \label{4}
    \end{figure}

As soon as the intrapocket pairing is also incorporated, the SC gap starts to oscillate around zero along all the pockets (Fig.~\ref{3}). In addition, the amplitude of oscillation in the SC gap has almost doubled for all the pockets. The gap along the larger hole pocket oscillates through zero twelve times for a complete rotation in the Brillouin zone, which is in contrast with four times oscillations through zero for smaller pockets. The amplitude of oscillation in the SC gap for smaller hole pocket and electron pocket is similar and nearly twice in comparison to the larger smaller pockets. Almost identifical gap functions for the smaller hole pocket and electron pocket results from the fact that these two pockets perfectly coincides with each other when translated by the momentum $\q = (\pi, \pi)$. The nature of gap in the DOS for the smaller hole pocket and the electron pocket indicates a pure $d$-wave SC gap while vanishing at $\omega = 0$. On the contrary, the gap in the DOS for the larger hole pocket despite having V-shape near $\omega = 0$, is finite and non zero.
Figure~\ref{4} shows the gap structure and behavior of DOS in the absence of the smaller hole pocket. We find that the nature of SC gap functions remains similar as in the case of three pockets except a reduction in the amplitude of oscillation.

\begin{figure}
   \includegraphics[scale=1.0, width=9cm]{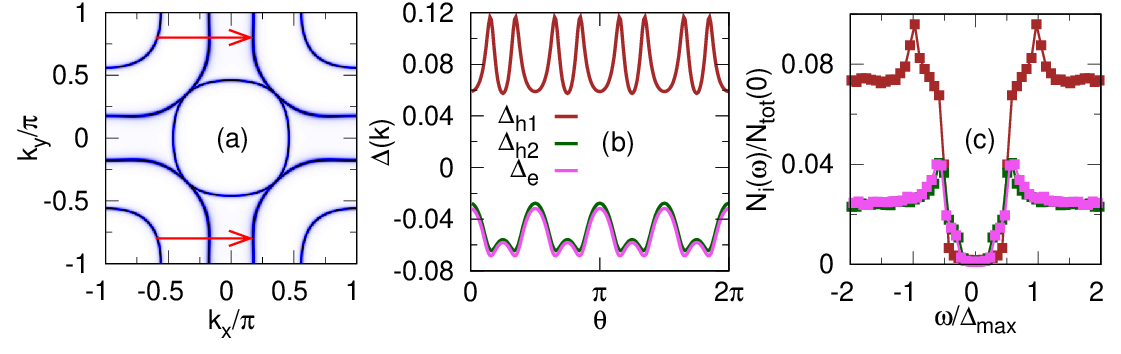}
    \vspace{-5mm}
   \caption{(a) Nesting vector, (b) SC gap structure, and (c) DOS for the self-consistently obtained solutions for pairing momentum $\Q \sim$ ($\pi$, 0). No self-consistent solution is obtained if the intraband pairing is also considered. }
    \label{5}
    \end{figure}

Next, we consider another widely debated pairing associated with the nesting vector $\Q \sim (\pi, 0$), which connects the nearly straight arms of smaller and larger hole pockets along the horizontal direction (Fig.~\ref{5}(a)). We could obtain a self-consistent solution only when the interband pairing is considered. We explored all types possibilities$-$including pure sign-changing $s$ wave, pure $d$-wave as well as mixed state of $s$- and $d$-waves. The gap functions along various pockets show oscillation with significant amplitudes, though the nature of the gap remains to be $s^{\pm}$ (Fig.~\ref{5}(b)). Although overall period of oscillation is $\pi/2$ but there is another maxima between the two maxima associated with four-fold rotation symmetry for the larger hole pocket.  Because of non-negligible oscillation in the gap parameter, the DOS shows a mixed character involving $s$- and $d$-wave away from the low-energy region. On the other hand, it has clear $s$-wave like behavior in the low-energy region (Fig.~\ref{5}(c)). It may be noted that, in the absence of the smaller hole pocket, the pairing associated with magnetic fluctuations with $\Q = (\pi, 0)$ cannot be supported.

\begin{figure}
   \includegraphics[scale=1.0, width=9cm]{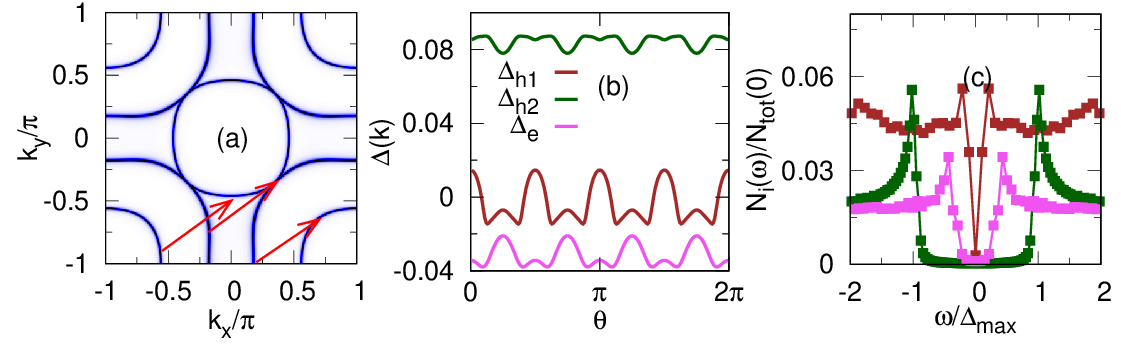}
    \vspace{-5mm}
   \caption{(a) Nesting vector, (b) SC gap structure, and (c) DOS for the self-consistently obtained solutions for the pairing momentum $\Q \sim$ ($0.75\pi$, $0.75\pi$) when only the interband pairing is considered. The SC gap along the larger hole pocket is $d$-wave like where it sign changing $s$-wave like for the smaller hole pocket and the electron pocket.}
    \label{6}
    \end{figure}
\begin{figure}
   \includegraphics[scale=1.0, width=9cm]{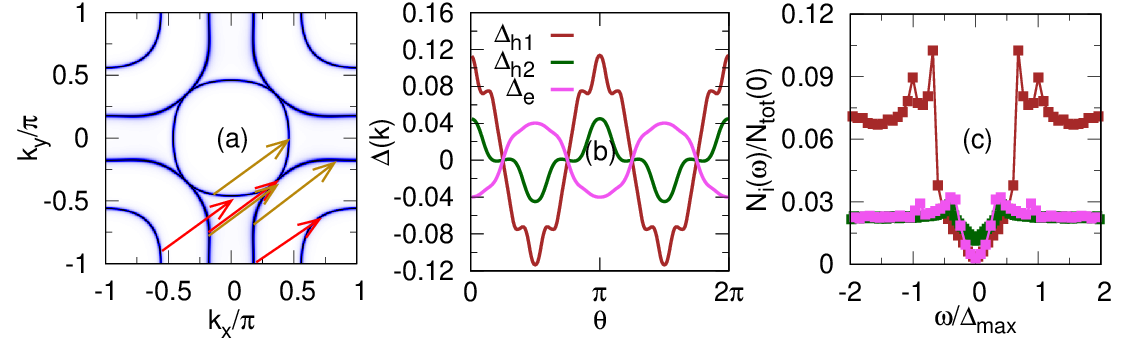}
    \vspace{-5mm}
   \caption{(a) Nesting vector, (b) SC gap structure, and (c) DOS for the self-consistently obtained solutions for pairing momentum $\Q \sim$ ($0.75\pi$, $0.75\pi$) when both the interband and intraband  pairings are incorporated simultaneously. All the SC gaps have similarity to a $d$-wave state while the gap along the smaller hole pocket differs the most as reflected through the corresponding DOS. }
    \label{7}
    \end{figure}

\begin{figure}
   \includegraphics[scale=1.0, width=9cm]{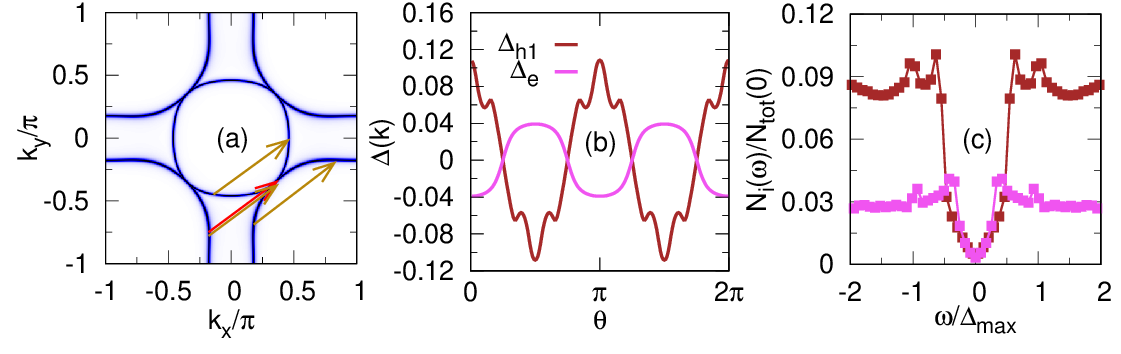}
    \vspace{-5mm}
   \caption{a) Nesting vector, (b) SC gap structure, and (c) DOS for the self-consistently obtained solutions for pairing momentum $\Q \sim$ ($0.75\pi$, $0.75\pi$)  when both interband and interband pairings are  considered while the smaller hole pocket is ignored~\cite{wang2}. The gaps along both the hole pocket and the electron pocket have $d$-wave like solutions.}
    \label{8}
    \end{figure}

\begin{figure}
   \includegraphics[scale=1.0, width=9cm]{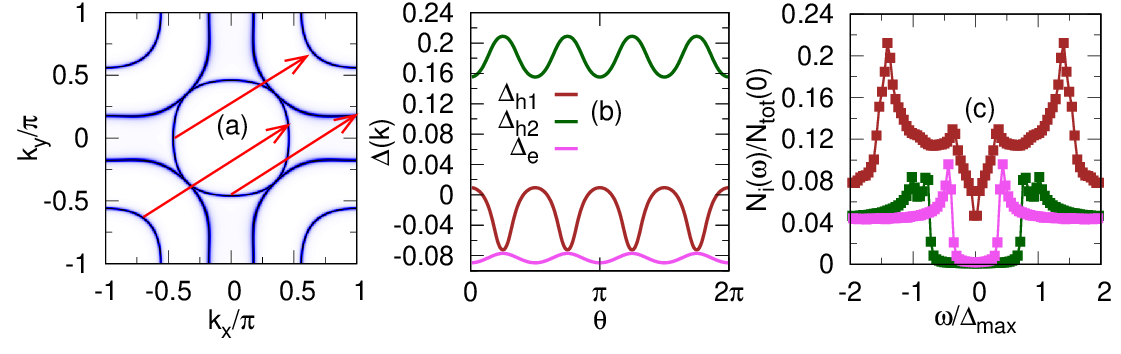}
    \vspace{-8mm}
   \caption{(a) Nesting vector, (b) SC gap structure, and (c) DOS for the self-consistently obtained solutions for pairing momentum $\Q \sim$ ($\pi$, $\pi$) when interband pairing is considered. Along the smaller hole pocket and electron pocket, the gaps are oscillating with opposite signs. Along the larger hole pocket, a significant oscillation in the gap is noticed and the gap also take negative value in a small region of the Fermi surface.}
    \label{9}
    \end{figure}

\begin{figure}
   \includegraphics[scale=1.0, width=9cm]{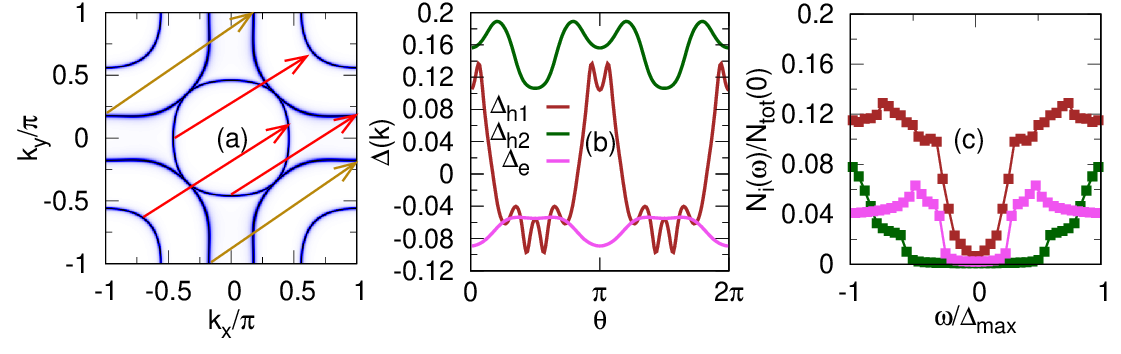}
    \vspace{-8mm}
   \caption{(a) Nesting vector, (b) SC gap structure, and (c) DOS for the self-consistently obtained solutions for pairing momentum $\Q \sim$ ($\pi$, $\pi$) when both the interband and intraband  pairings are incorporated simultaneously. The larger hole pocket has $d$-wave like gap structure whereas the remaining Fermi pockets have $s$-wave like gap with sign opposite. }
    \label{10}
    \end{figure}

\begin{figure}
   \includegraphics[scale=1.0, width=9cm]{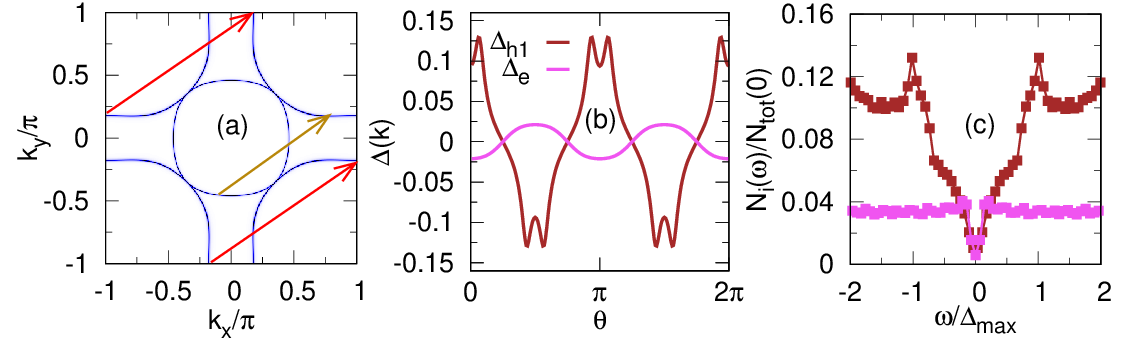}
    \vspace{-8mm}
   \caption{(a) Nesting vector, (b) SC gap structure, and (c) DOS for the self-consistently obtained solutions for pairing momentum $\Q \sim$ ($\pi$, $\pi$) when both the interband and interband pairings are considered whereas the smaller hole pocket is ignored~\cite{wang2}. The gaps along both the hole pocket and the electron pocket have $d$-wave like structure.}
    \label{11}
    \end{figure}

The next larger and possible pairing wavevector in magnitude is $\Q \sim$ (0.75$\pi$, 0.75$\pi$) as suggested by the first-principle band structure calculation~\cite{yang}. It may be noted that this possible pairing wavevector does not correspond to a commensurate magnetic correlation structure. Figure~\ref{6}(a) shows some of the feasible nesting vectors giving rise to such magnetic correlations when only the interband pairing is allowed. The gap size is smaller along the electron and larger hole pocket while it is the largest along the smaller hole pocket. Moreover, the SC gap functions on the electron and smaller hole pocket are opposite in sign, whereas the mean value of the gap along the larger hole pocket is nearly zero. The latter is also reflected in the DOS. The DOS along the larger hole pocket is $V$-shaped. As expected, the DOS along the remaining pockets show $s$-wave character. Upon inclusion of the intraband pairing, the $d$-wave like SC gap is found along all the pockets (Fig.~\ref{7}(b)). In particular, the amplitude of oscillation is now the largest along the larger hole pocket. In fact, the amplitude of the SC gap oscillation is largest of all the cases considered till now. Secondly, the gap along smaller hole pocket becomes nearly flat in smaller sections. The consequence of these gap features can be clearly seen in the DOS. Almost a V-shape gap in the DOS exists for all the pockets (Fig.~\ref{7}(c)). The gap full opens for the larger hole-pocket and electron pocket, but it does not for the smaller hole pocket.

Finally, we discuss the scenario, when the pairing is mediated by the $\Q \sim$ ($\pi, \pi$)-type magnetic correlations. In this case, the nesting between the Fermi pockets responsible for the generation of short-range magnetic fluctuations can be the intrapocket nesting because of the larger hole pocket as in the case of high-$T_c$ cuprates. A significant contribution can also come from the interpocket nesting between the electron and smaller hole pocket. We once again start first by considering only the interband pairing. Figure~\ref{9} shows the gap structure. The average gap size along the larger hole pocket is small however with a large amplitude. On the other hand, the SC gap size along the smaller hole pocket is very large and opposite in sign when compared to the larger hole pocket as well as the electron pocket, i.e., $\Delta^a_{h1} <  \Delta^a_e <\Delta^a_{h2}$. This results from a good nesting between the smaller hole pocket and electron pocket, and a relatively poor but non negligible nesting between the smaller and larger hole pocket. However, the nesting is very weak between the electron pocket and the larger hole pockets. The SC gap along the larger hole pocket is not strictly negative, i. e., it crosses zero. It crosses in such a way the maxima of the SC gap is in the vicinity of zero. As a result, there is finite DOS at $\omega = 0$, i. e., the gap in the DOS is not fully open while it has a distinct $V$-shape. On the other hand, the DOS has a clear $s$-wave signature along the remaining pockets.

Upon incorporating the intraband pairing, which may be mostly contributed by the largest hole pocket as in the case of high $T_c$ cuprates, we find that a mixed state consisting of $s$-wave along the electron and smaller hole pockets whereas a $d$-wave like gap along the larger hole pocket is stabilized (Fig.~\ref{10}). It may be worthwhile to note that the average $s$-wave gap as well as amplitude of $d$-wave is the largest for a given set of parameters in comparison to all the cases considered till now in the current work. It indicates the robustness of the mixed SC state with pairing wavevector $\Q \sim$ ($\pi, \pi$). The nature of SC state further substantiated by the nature of DOS which displays V-shape along the larger hole pocket especially in the low-energy region and $s$-wave character for all the other pockets. Interestingly, in the absence of smaller hole pocket, only pure $d$-wave SC state is stabilized along both the remaining pockets. Figure~\ref{11} shows the results obtained in the absence of smaller hole pocket. The gap and DOS structures remain more or less similar except the magnitude.

\section{Conclusions}
To conclude, we have examined the stability of pure and mixed superconducting state within an effective three-band model with a focus on dependence of stability on different types of pairing wavevectors associated with the magnetic correlations. Our findings suggest that a mixed superconducting state for the pairing momenta $\Q \sim$($\pi, \pi$), consisting of sign changing $s$-wave like gap across the electron and smaller hole pockets while a $d$-wave like gap along the larger hole pocket, is likely to be more stable in comparison to a pure sign-changing $s$-wave or $d$-wave. In all the cases of possible pairing wavevectors explored here, we find that there is significant oscillation in the amplitude of  oscillation for the $s$-wave gap functions originating from the presence of multiple Fermi pockets which are large in size. One important consequence of such large oscillations in the amplitude is a $d$-wave like behavior in the high-energy region of density of states and $s$-wave like character in the low-energy region. Our approach focused largely on the role of geometrical shape of the Fermi pockets on pairing mechanism and ignored the details of orbital content of these pockets. However, we do partially incorporate the effect of orbital degree of freedom, for instance, important details such as bilayer splitting of bands is incorporated into our effective band model. The role of orbital-weight distribution along the Fermi surface could be taken into account, which is unlikely to modify our main conclusions, particularly when the bands have nearly equal contribution from both the orbitals.

\end{document}